# An Electroencephalography connectome predictive model of major depressive disorder severity

Aya Kabbara[1,2], Gabriel Robert[3,4,5], Mohamad Khalil[6,7], Marc Verin[5], Pascal Benquet[8], Mahmoud Hassan[2,9]

[1] Lebanese Association for Scientific Research, Tripoli, Lebanon

[2] MINDig, F-35000 Rennes, France

[3] Academic department of Psychiatry, Centre Hospitalier Guillaume Régnier, France

[4] Empenn, U1228, IRISA, UMR 6074, Rennes, France

[5] Comportement et noyaux gris centraux, EA 4712, CHU Rennes, Université de Rennes 1, Rennes 35000, France.

[6] Azm Center for Research in Biotechnology and Its Applications, EDST, Tripoli, Lebanon.

[7] CRSI research center, Faculty of Engineering, Lebanese University, Beirut, Lebanon

[8] Univ Rennes, Inserm, LTSI - U1099, F-35000 Rennes, France

[9] School of Science and Engineering, Reykjavik University, Reykjavik, Iceland.
Corresponding author: Mahmoud Hassan, mahmoud.hassan.work@gmail.com

# Abstract

Emerging evidence showed that major depressive disorder (MDD) is associated with disruptions of brain structural and functional networks, rather than impairment of isolated brain region. Thus, connectome-based models capable of predicting the depression severity at the individual level can be clinically useful. Here, we applied a machine-learning approach to predict the severity of depression using resting-state networks derived from source-reconstructed Electroencephalography (EEG) signals. Using regression models and three independent EEG datasets (N=328), we tested whether resting state functional connectivity could predict individual depression score. On the first dataset, results showed that individuals scores could be reasonably predicted ($r$=0.61, $p$=4 x $10^{-18}$) using intrinsic functional connectivity in the EEG alpha band (8-13 Hz). In particular, the brain regions which contributed the most to the predictive network belong to the default mode network. We further tested the predictive potential of the established model by conducting two external validations on (N1=53, N2=154). Results showed high significant correlations between the predicted and the measured depression scale scores ($r1$= 0.49, $r2$=0.37, $p$<0.001). These findings lay the foundation for developing a generalizable and scientifically interpretable EEG network-based markers that can ultimately support clinicians in a biologically-based characterization of MDD.

# Introduction

Major Depressive Disorder (MDD) is one of the most common psychiatric disorders, mainly characterized by anhedonia and disturbed mood causing psychological, social, functional and economic consequences [1]. Due to its increasing prevalence, chronicity, recurrence and degraded quality of life, MDD is now considered as a public health problem [2,3]. Currently, there are no biological signature of MDD, most probably because of its heterogeneity, and therefore prognosis (including better treatment response) and even diagnostic can be, sometimes, challenging. Researchers in mental illness advocate for a more biologically based framework to diagnose and treat these disorders, including depression [4,5]. At the cerebral level, most recent advances have moved from localized cerebral area disruptions to more network-based measures of mental disorders [6].

Emerging evidence across functional studies consistently points at disruptions of MDD brain networks both in resting state [7–19] and during task-based connectivity. Specifically, functional abnormalities in network topology [11,20,21], modularity [10,22], and efficiency [8,11,15,16,23] have been detected between MDD and healthy controls. Locally, these alterations are predominantly observed in the Default-Mode Network (DMN) regions, including the posterior cingulate [23,24], hippocampus [7,8,23,25], parahippocampal gyrus [7,20,23], precuneus [23,26] superior parietal lobule [18,23,26], and the executive network (ECN) including dorsolateral prefrontal cortex [8,9,14,25] and the anterior cingulate cortex [8–10,13,15–18,20,21,25]. Importantly, DMN abnormal connectivity patterns is thought to be related with the cognitive vulnerability [27,28] and negative self-referential thoughts of MDD patients [29,30]

While this previous work has enabled a step toward a better understanding of the underlying pathophysiology of MDD, their translational potential into clinical use is hampered by the cost-effectiveness and availability of MRI. Electroencephalography (EEG) combined with the use of easy-to-implement measures of the brain network have been used to both binary classify [31–33] and to predict treatment outcomes [34,35] in MDD. However, binary classification does not answer the question of MDD heterogeneity nor testing the possible continuum between normal sadness and pathological major depression. Also, the quantification of the MDD symptoms is required to fully address the problem of prognosis. Very recently, resting state global connectivity measures predict both depressive and anxiety symptoms [36–38]. However, machine-learning/multivariate based connectivity measures are of high dimension and therefore often hampered by overfitting biases [39]. Proper analyses require large sample size and independent sample to validate the trained algorithm, which has not been the case in the field of EEG and affective disorders.

In this paper, we aim to establish a model that predicts depression severity based on resting-state EEG cortical networks. To tackle the problem of generalizability, three independent large datasets (N=328) were used. Based on the prior findings, we hypothesized that the key regions that might contribute to the predictive model comprised DMN and ECN regions, mainly the anterior cingulate cortex, the prefrontal cortex, and the parahippocampal regions. We also expected that individual differences in MDD would be predicted by characteristics of alpha band as many studies suggest its relation to the pathologic characteristics of MDD [40–42]. EEG signals collected from 121 subjects (76 healthy controls and 45 MDD patients) were used to derive predictive models for depression severity. Brain networks were reconstructed using EEG source connectivity method [43]. Then, the Connectome-based Predictive Modeling (CPM) [44] was applied in order to establish the relationship between functional brain networks and depression severity in support vector regression model. The external validation was performed using two other independent EEG datasets, where the first includes 24 MDD and 29 healthy controls, and the second (N=154) -with no MDD patients- shows an ultimate perspective of predicting a possible continuum between normal sadness and pathological major depression.

# Materials and methods

The full pipeline of this study is summarized in Figure 1.

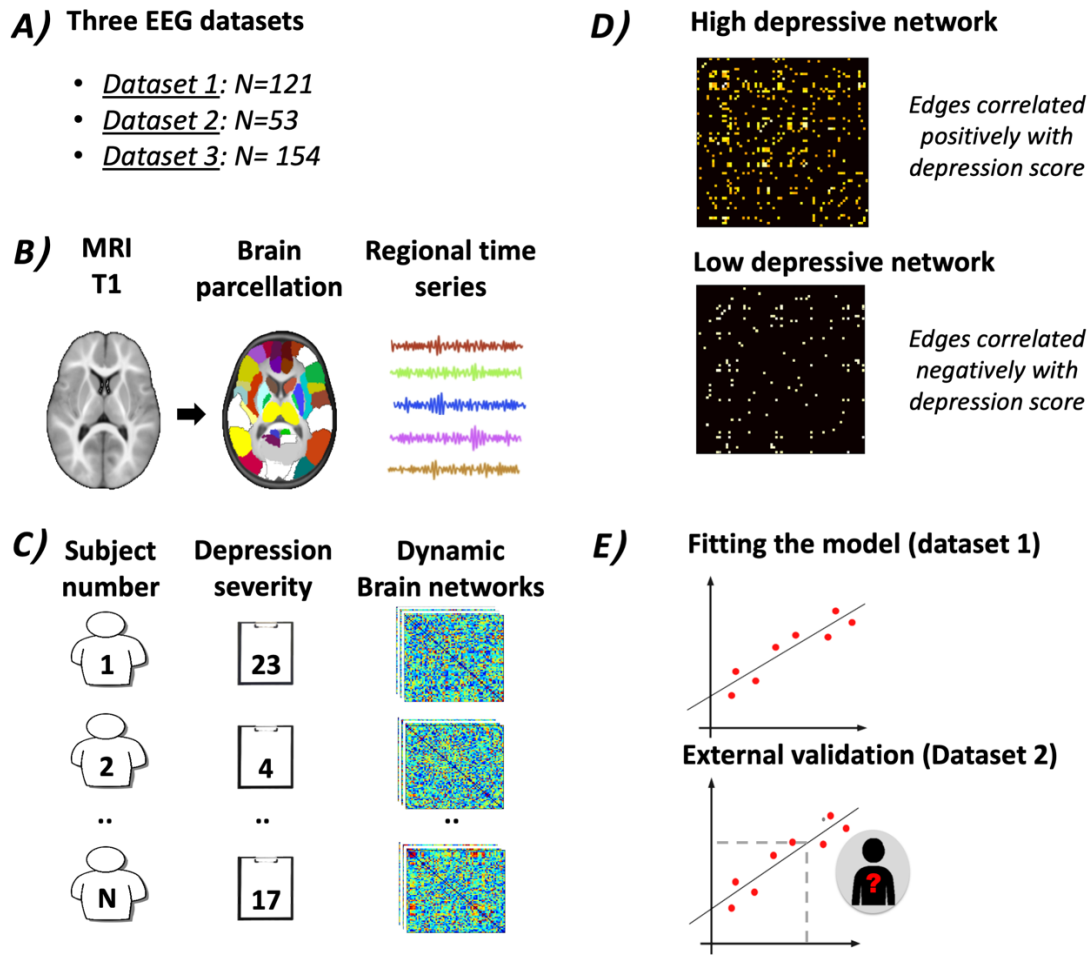

**Figure 1. The full pipeline of our study. (A) EEG signals were acquired from three datasets: 1- N= 121 subjects (76 healthy controls and 45 MDD patients), 2- N=53 subjects (24 MDD patients and 29 healthy controls) 3- N=154 subjects. (B) A template Magnetic Resonance Imaging (MRI) was segmented into 68 regions of interest (ROIs) by the means of Desikan Killiany atlas. Then, the regional time series of each subject were reconstructed using the weighted minimum norm estimate inverse solution (WMNE). (C) The inputs of the connectome-based predictive modeling (CPM) are the connectivity matrices and the depression score of each subject (BDI (i.e. dataset1), Hamilton (i.e. dataset2, dataset 3)). The brain networks of each subject were obtained by computing the phase locking value between the regional time series. (D) The training step: Across all subjects, each edge is correlated to the BDI score. Then, the algorithm selects the most important edges using significance testing (p<0.01). Two matrices are thus resulted: The first corresponds to the network positively correlated with BDI and the second represents the network negatively correlated with BDI. (E) A predictive model is built assuming a linear relationship between BDI and the sum of edge weights of**

**the matrices obtained in the previous step. This model is then applied to predict the depression severity of subjects from the second and third dataset.**

# First dataset

## Participants

Forty-five MDD patients and seventy-five healthy controls have participated in the current study. The EEG database is publicly available at http://bit.ly/2rzY6ZY and was used in a recent previous study [45]. The study was approved by the ethical committee of Arizona University and experiments were in accordance with relevant ethical guidelines. Prior to acquisition, all participants provided written informed consent. Participants were recruited from introductory psychology classes based on mass survey scores of the Beck Depression Inventory (BDI). Recruitment criteria included: (1) age 18–25, (2) no history of head trauma or seizures, and (3) no current psychoactive medication use. Control participants ($N = 76$) have stable low BDI (<7) between mass survey and preliminary assessment, no self-reported history of MDD, and no self-reported symptoms indicating the possibility of an Axis 1 disorder as indicated by computerized self-report completion of the Electronic Mini International Neuropsychological Interview (eMINI: Medical Outcome Systems, Jacksonville, FL, USA). Depressed participants have a stable high BDI (>13). Participant demographics are reported in Table 1.

**Table 1. Demographics of the 121 participants. BDI_Anh= anhedonia subscale of BDI; BDI_Mel= Melancholia subscale of BDI; TAI= trait anxiety inventory**

|  | HC | MDD | p-value |
|---|---|---|---|
| Cases (N) | 76 | 45 | NaN |
| Gender (M/F) | 35/40 | 12/33 | 0.02 |
| Age (years) | 18.9 ± 0.6 | 19 ± 0.55 | 0.15 |
| **Symptom scores** | | | |
| BDI | 1.6 ± 0.75 | 22 ± 2.7 | 1E-20 |
| BDI_Anh | 0.16 ± 0.2 | 4 ± 0.8 | 1E-22 |
| BDI_Mel | 0.81 ± 0.4 | 6 ± 1 | 1E-17 |
| TAI | 30.87 ± 2.8 | 56 ± 4.6 | 1E-20 |

# EEG acquisition and pre-processing

Participants were asked to stay relaxed for five minutes while EEG signals were recorded. Signals were acquired using 64 Ag/AgCl EEG electrodes (Synamps system) positioned according to the standard 10–20 system montage, two electro-oculogram electrodes (EOG) for horizontal and vertical movements. Signals were sampled at 500

Hz, bandpass filtered between 0.5 and 100 Hz. All electrodes' impedances were kept below 10 kΩ.

Because of contamination of the EEG signals by various types of artifacts, EEG (pre)processing was applied following the same steps proposed in several previous studies dealing with EEG resting-state data [46,47]. These steps are summarized as follow: 1) Identification of bad channels providing signals that are either completely flat or contaminated by movement artifacts. This was performed by visual inspection, complemented by the power spectral density, 2) Interpolation of identified bad channels in Brainstorm by using neighboring electrodes within a 5-cm radius. 3) Segmentation into 40-s length epochs and only epochs with voltage fluctuations between +100 μV and −100 μV were retained. For each participant, four artifact-free epochs of 40-s lengths were selected. This epoch length guarantees a good compromise between the needed temporal resolution and the results reproducibility as demonstrated in [46]. Due to poor signal quality, EEGs from three HC subjects were excluded from the analysis.

## Second dataset

### Participants

A total of 24 subjects (30.88 ± 10.37 years, female=11) diagnosed with major depressive disorder, and 29 healthy controls (31.45 ± 9.15 years, female=9) were recruited for the study. Patients with MDD were recruited from Lanzhou University Second Hospital, Gansu, China, diagnosed and recommended by professional psychiatrists while healthy subjects were recruited by posters. The study was approved by the Ethics Committee of the Second Affiliated Hospital of Lanzhou University. All procedures performed in the study were in accordance with the ethical guideline of the national research committee of Lanzhou University. Written informed consent was obtained from all subjects before the experiment began. All MDD patients received a structured MiniInternational Neuropsychiatric Interview (MINI) that meets the diagnostic criteria for major depression of Diagnostic and Statistical Manual of Mental Disorders (DSM) based on the DSM. The dataset is publicly available in. Depression symptoms were identified with Hamilton Depression Scale.

## EEG acquisition and pre-processing

During experiment, participants were required to relax, stay awake and try avoiding movements while 5 minutes eye-closed resting state EEG were recorded in a sound-proof room. 128-channel HydroCel Geodesic Sensor Net (Electrical Geodesics Inc., Oregon Eugene, USA) was used and electrode signals were referenced to Cz. During acquisition, each electrode kept an impedance below 50 kΩ. EEG preprocessing was performed in the same way dealt with the first dataset, leading to four artifact-free epochs of 40-s lengths for each participant.

# Third dataset

## Participants

154 young healthy participants (N=154, 25.1±3.1 years, range 20–35 years, 45 female) were provided by the Mind-Brain-Body dataset[48] publicly available in **http://fcon_1000.projects.nitrc.org/indi/retro/MPI_LEMON.html**. The study was carried out in accordance with the Declaration of Helsinki and the study protocol was approved by the ethics committee at the medical faculty of the University of Leipzig (reference number 154/13-ff). All included participants provided written informed consent before any data acquisition. The dataset originally included also 75 elderly subjects who were removed from the analysis to avoid the impact of age on the external validity of the model. Depression symptoms were identified with Hamilton Depression Scale.

## EEG acquisition and pre-processing

Participants were instructed to be awake with their eyes open and fixate on a low-contrast fixation cross on grey background. For each subject, 16-min resting state EEG was recorded with a 'BrainAmp plus' amplifier EEG using both 62-channel (61 scalp electrodes plus 1 electrode recording the VEOG below the right eye) and active ActiCAP electrodes (Brain Products GmbH, Gilching, Germany) positioned according to the international standard 10–20 extended localization system. Electrodes were referenced to FCz, the ground was located at the sternum and electrode impedance was kept below 5 KΩ. EEG were bandpass filtered between 0.015 Hz and 1 kHz and sampled at 2500 Hz.

Data were provided pre-processed, after passing through a pipeline that removed artefactual segments, identified faulty recording channels, and regressed out artefacts which appear as independent components in an Independent Component Analysis (ICA) decomposition with clear artefactual temporal signatures (such as eye blinks or cardiac interference).

## Brain networks construction

Brain networks were reconstructed using the "EEG source connectivity" method [43] following two main steps: i) estimate the cortical sources and reconstruct their temporal dynamics by solving the inverse problem, and ii) measure the functional connectivity between the reconstructed regional time-series.

In brief, after co-registering EEGs and MRI template (ICBM152), a realistic head model was built using OpenMEEG tool [49]. Then, the cortical surface was parcellated into 68 regions of interest by the means of Desikan-Killiany atlas [50]. As an inverse solution, the Weighted Minimum Norm Estimate (wMNE) algorithm was selected to estimate the regional time series. Afterwards, we filtered the reconstructed regional time series in the EEG frequency bands (delta: 1–4 Hz; theta: 4–8 Hz; alpha: 8–13 Hz; beta: 13–30 Hz and gamma: 30-45 Hz). In order to finally reconstruct the functional network, the Phase Locking Value (PLV) was assessed between the regional time series. This process applied yields an undirected and weighted connectivity matrix for each subject. We used the wMNE/PLV combination to reconstruct the dynamic networks, as it is widely used in the context of EEG source-space networks at rest [46,51] and it is supported by several model-based and real data-based comparative studies [52].

## Connectome Predictive Modeling

The CPM [44] is a recently developed method for identifying and modeling a brain network associated with a variable of interest, the BDI in our case. CPM was previously employed in a number of studies to predict network alteration in several brain disorders such as anxiety related illnesses [10] and sleep disorders as well in some other conditions such as personality traits and creativity [10,53].

# Constructing the predictive networks

We established the predictive model using the first EEG dataset. For each subject, the connectivity matrices were averaged across time samples and epochs. Then, each edge in the connectivity matrix was correlated with the BDI scores across all participants using Pearson's correlation. Afterwards, only edges that show significant correlations with BDI ($p_{FDR}$< 0.01) were retained. The correlation values were separated into a positive tail (i.e. edges correlated positively with BDI scores) and a negative tail (i.e. edges correlated negatively with BDI scores). Therefore, this step will result in the reconstruction of two networks: high depressive network (including connections in the positive tail) and low depressive network (including connections in the positive tail). Following this, we extracted the edge strength in both positive and negative tails in order to correlate it with BDI, as proposed by [44]. The edge strength (summed index) is obtained by summing the Z scores of all connections in the positive and negative tails.

# 5-fold cross validation (internal validation)

To investigate whether the edge strength can predict the BDI scores in novel subjects, a 5-fold cross validation was used to evaluate the prediction performance. Particularly, it was reported that K-fold cross-validation procedure provides more accurate model evaluation compared to holdout method and leave-one-out procedure. Here, 5-fold cross-validation method was performed by splitting the 121 subjects into 5 subsets of data (also known as folds). Each regression model is trained on all but one of the subsets by identifying the predictive networks, and computing the corresponding edge strengths from the training set used to construct a Support Vector Regression model (SVR) that respectively relate positive and negative edge strengths to BDI scores. Then, the trained model is tested on the subset that was not used for training. This process is repeated 5 times, with a different subset reserved for evaluation (and excluded from training) each time. The performance assessment measures are calculated for each time and the final performance is averaged over the five folds. The correlation between the actual BDI and the predicted scores was evaluated using Pearson's correlation. The 5-fold cross validation was repeated for 100 iterations, and the resulted performance presents the average performance across all iterations. Finally, in order to check if the obtained correlation is significantly better than expected by chance, we performed a permutation test by permuting the BDI scores of all participants for 1000 times. This yields in a null distribution of correlation values.

## External validation

The derived high and low depressive networks were used as predictive networks for the second and third independent dataset. In brief, we computed the edge strength within high and low depressive networks established as predictive models using the first dataset. This feature was then correlated with the Hamilton depression scale score in these novel samples. In addition, we computed for each subject, a summed edge index that combines both positive and negative as proposed by [54]. This latter is obtained by summing the Z-scores of all connections of the positive and subtracting those obtained from the negative tails.

# Results

The full pipeline of the study is illustrated in Figure 1. Resting-state EEG data were obtained from three independent datasets. After EEG pre-processing, the cortical brain networks were assessed for each subject using EEG source connectivity method [55]. We used the wMNE followed by phase-couplings between 68 brain regions defined using Desikan-Killiany atlas [50]. Then, we conducted a whole-brain search for the predictive connectome features using the first dataset (N=121) served for the training and internal validation of the predictive model. The connectome-based model identifies the predictive brain connections (i.e edges) as those that show significant positive or negative correlations with the Z-score (p<0.01). A positive (negative) correlation means that the correspondent connection predicts high (low) depression severity. Afterwards, we performed external validation using the second and third datasets (N=197).

## Demographics data

No statistical difference in age ($p$ = 0.15) was depicted between MDD and HC. Results show the existence of a statistical difference in gender ($p$ = 0.02) as computed using chi-square test. MDD patients and HC differ in all the recorded psychiatric scores ($p$<0.0001) as computed using the Wilcoxon ranksum test. For the second dataset, there were no significant between-group differences in age ($p$ = 0.832), or gender ($p$ = 0.269).

Statistical differences in the self-reported PHQ-9, and the Generalized Anxiety Disorder-7 (GAD7), the Hamilton depressive scores were obtained ($p < 0.001$).

## Predictive networks

In this study, we are more interested in quantifying MDD heterogeneity in terms of brain connectivity rather than classifying binary groups. The binary classification will be very limited to deal with the MDD heterogeneity or testing the possible continuum between normal sadness and pathological major depression. Thus, we used the BDI scores that quantify MDD severity, to construct the predictive networks of depression using the first dataset. The high-depressive network denotes the network in which edges show positive significant correlations with BDI scores, and the low depressive network denotes the network in which edges show negative significant correlations with BDI scores. Gender was considered as a covariate in the model estimation. No significant networks were detected in delta, theta, beta and gamma bands. In alpha band, the high depressive network exhibited dense functional connections in predominantly prefrontal, insula and limbic lobes (Figure 2). More specifically, the implicated brain regions are: the caudal middle frontal gyrus (cMFG), insula (INS), parahippocampal (paraH), posterior cingulate (PCC), and rostral anterior cingulate (rACC). The low-depressive network showed edges within occipital, parietal and limbic lobes. Nodes are the lateral occipital gyrus (LOG), the superior parietal lobule (SPL) and the precunues (PCUN). Notably, the implicated regions in the predictive networks belong mainly to the default mode network (paraH, PCUN, PCC, SPL).

# Internal validation using dataset 1: 5-fold cross validation

We have evaluated the correlation between observed and predicted BDI score to assess the predictive performance of CPM in both models (high and low depressive networks). Figure 3.A illustrates that the predictions of the low-depressive network were significantly correlated with the observed BDI scores ($r$=0.37, $p$= 2.28E-05). Figure 4D shows the distribution of the correlation coefficients computed using the permuted BDI scores in the low-depressive network. The correlation value obtained using the observed BDI scores was considered significant when higher than the mean±2std.

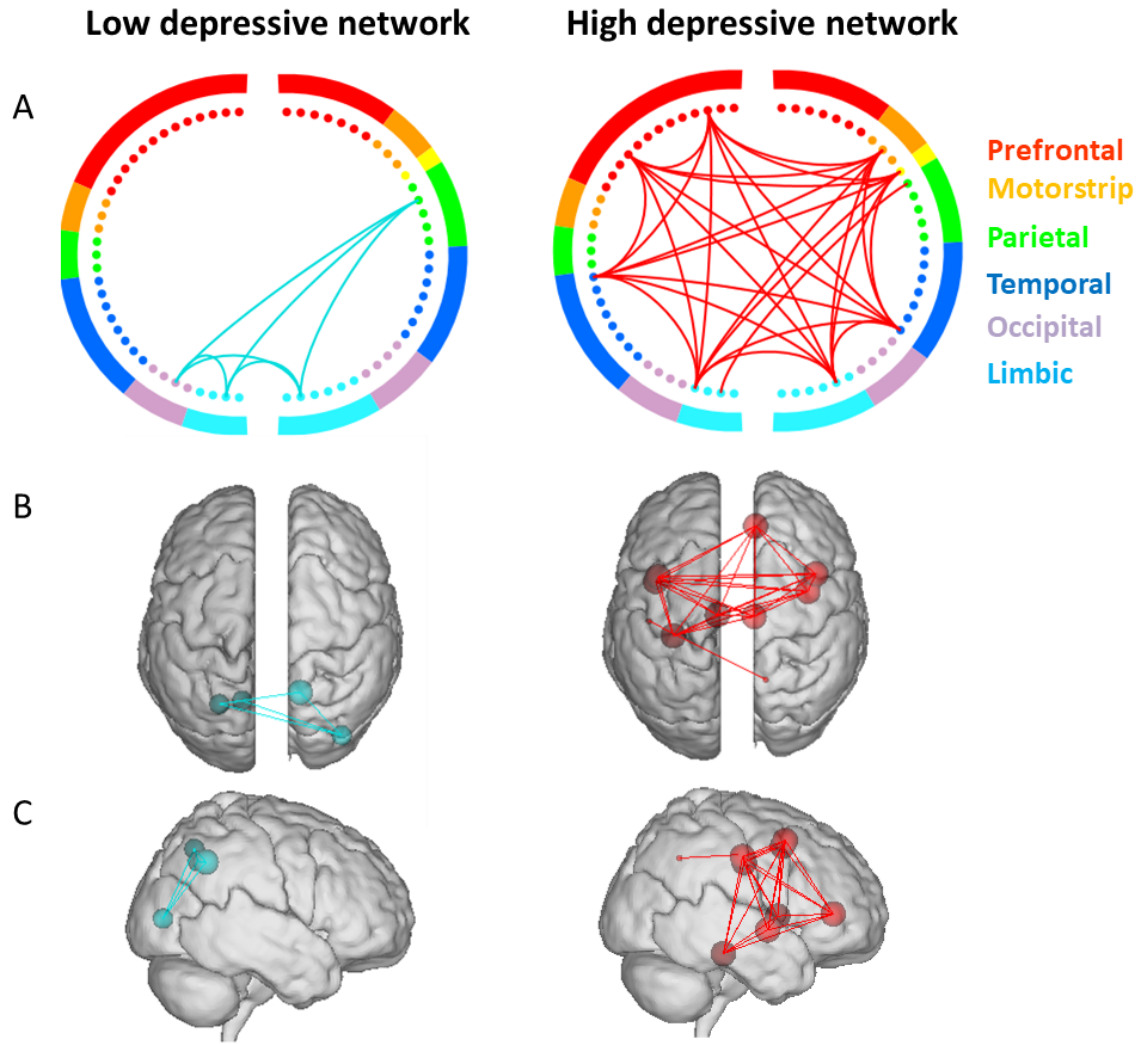

**Figure 2. Depictions of the high- and low-depressive networks. Circle plots (A), glass brains – top view(B) , glass brain- right view were obtained by keeping the signififcant edges (p<0.01, FDR corrected). Colors within the circle plots correspond to lobes of the brain.**

In addition, a positive relationship was detected between the predicted score and the observed BDI score in the high-depressive network ($r$=0.5526, p = 4.9E-11) as illustrated in Figure 3B.).

We also computed the correlation between the predicted and the observed depressive scores using both high and low depressive networks (Figure 3C). This was performed by calculating the summed index also named as edge strength by summing the Z scores of all connections in the positive and negative tails (see materials and methods). Then, this index will serve as the input feature of the SVR model. Using both high and low depressive edges, a positive significant correlation of r=0.64, *p*=4.4E-18 was obtained.

All prediction results are significantly better than expected by chance when compared to the null distributions (Figure 3D, Figure 3E, Figure 3F).

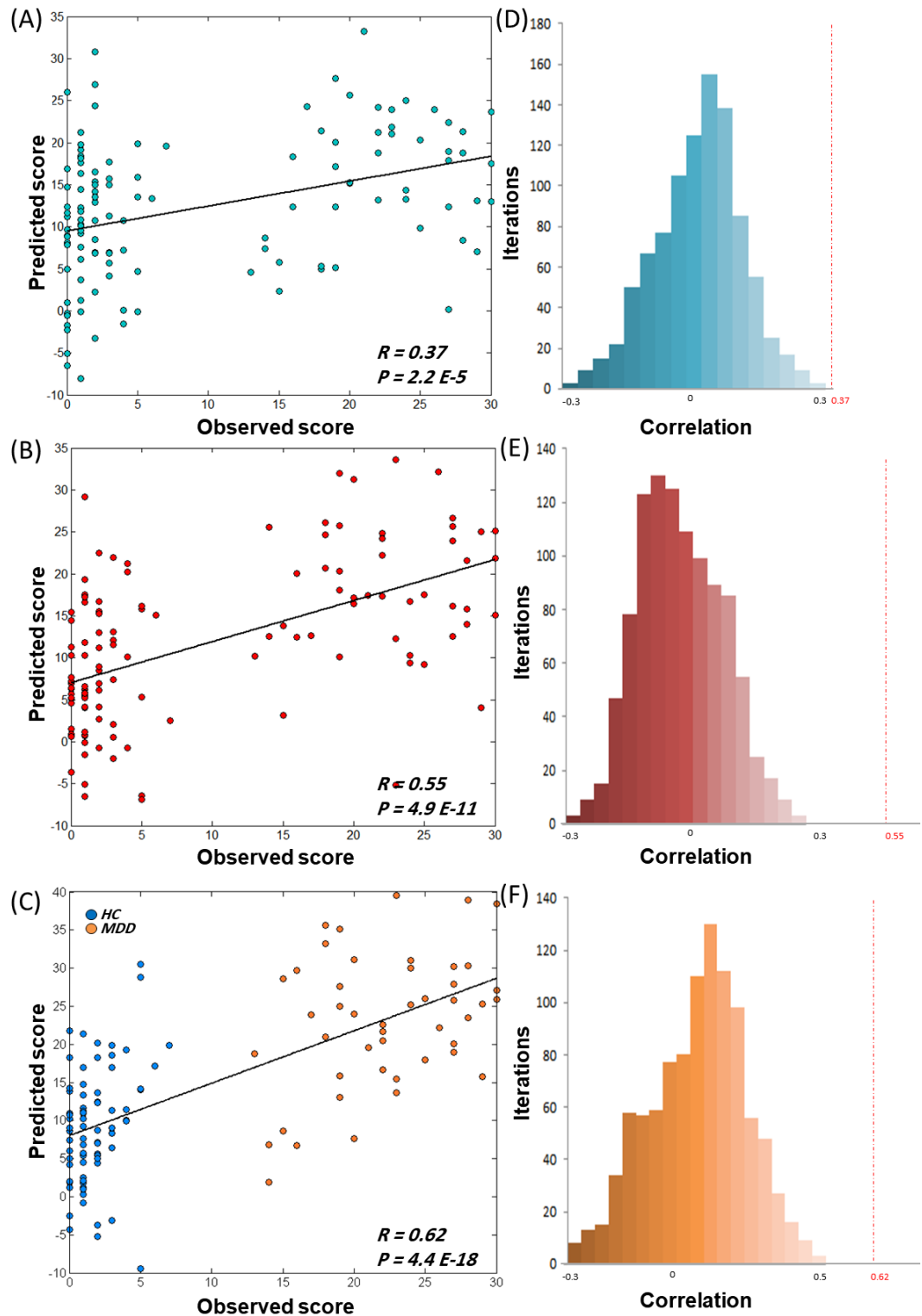

**Figure 3.** The relationship between the observed and the predicted scores using A) low depressive edges, B) high depressive edges, C) combined depressive edges (summed index). D)E)F)) The distribution of correlation values using permutation test on 1000 iterations computed, respectively, for the low depressive network, high depressive network and both networks. The dash line indicates the correlation value obtained using the predicted BDI scores.

# External validation on datasets 2 and 3

In order to test the generalizability of our results, we tested whether the edge strength of connectivity (sum of the correlation coefficients) within the high- and low-depressive networks defined in dataset1 could predict the depression severity in two other independent datasets (N=197). The edge strength was computed for the brain networks computed in alpha band where the model was derived.

For the second dataset (N=53), the depression scores were predicted using both high ($r$=0.36, $p$=0.007) and low predictive models ($r$=0.4, $p$=0.003) as shown in Figure 4A, B. When combining the high and low predictive connections (Figure 4C), the correlation between the observed and predicted scores was also significant ($r$=0.49, $p$=1.68E-4).

For the third dataset (N=154), our findings reveal that the SVR model was able to predict the depression scores for the high-depressive network ($r$=0.37, $p$=7 x $10^{-7}$), but not the low-depressive network ($r = -0.008$, $p = 0.9$) as reported in Figure 4E.

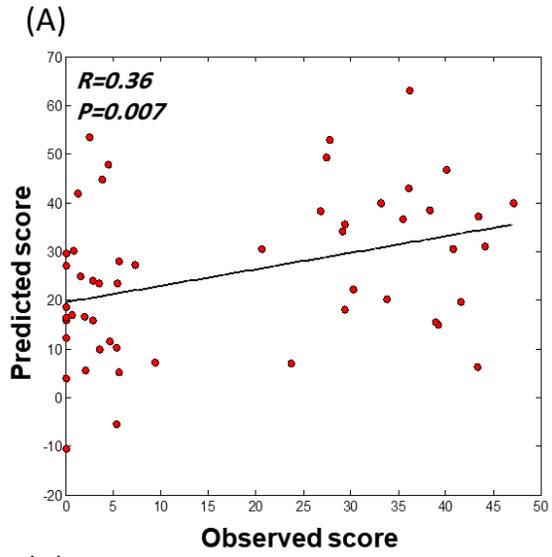
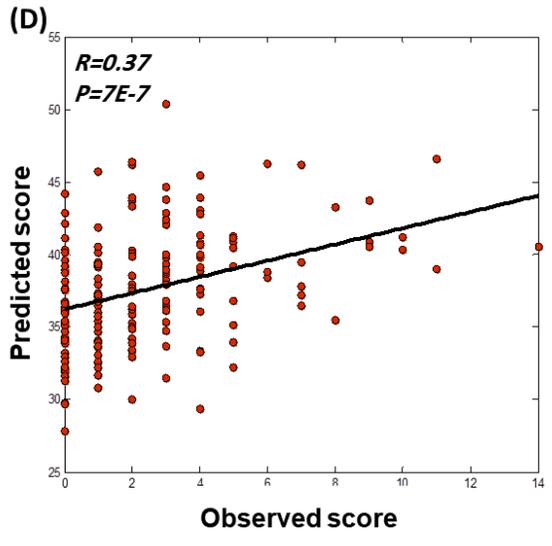
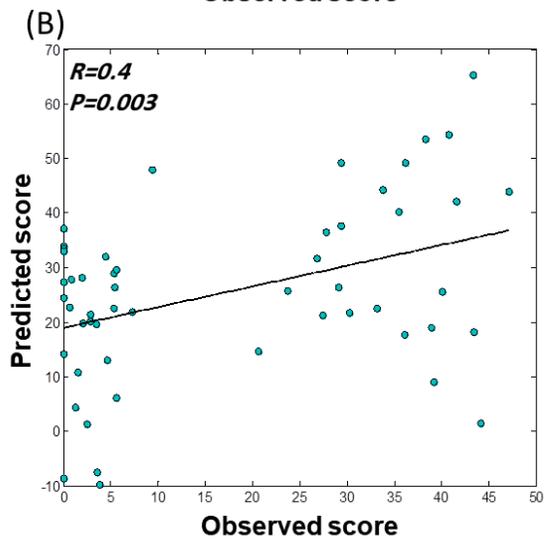
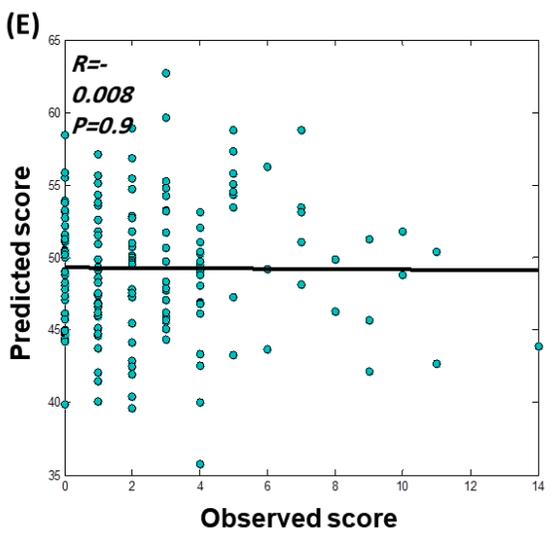
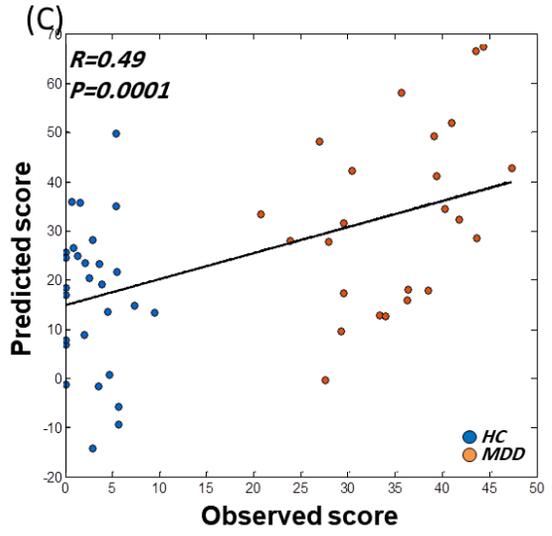

**Figure 4.** A) Correlation between the predicted and observed Hamilton scaling score of the second dataset using high-depressive network, B) The prediction results of the second dataset using low-depressive network, C) The prediction results of the second dataset using both high and low depressive network D) The prediction results of the third dataset using high depressive network, E) The prediction results of the third dataset using low depressive network.

# Discussion

The objective of this study is to establish a connectome-based model capable of predicting depression score at the individual level. Using three independent resting state EEG datasets, our findings proved that depression could be predicted using intrinsic functional connectivity. More specifically, the majority of the nodes that contributed the most to the predictive network belong to the DMN obtained in the alpha frequency band. Interestingly, generalizability of the model to predict novel and independent samples was demonstrated across different datasets as well as across different psychometric tests (BDI, Hamilton).

## Brain regions implicated in predictive network

Alterations in the paraH connectivity were previously reported [7,20,23]. As this region plays a major role in memory retrieval, it is supposed that such alteration is associated with the recall or the imagination of negative memories and events in MDD patients [56]. Consistent with our findings, previous studies have reported higher activation of paraH in depressed subjects compared to healthy group during rumination (moments) [57].

In addition, increased blood flow in left and right PCC regions were associated with MDD [58]. Using EEG source connectivity combined with graph theory, a recent EEG study shows that MDD brain network is characterized by increased efficiency in PCC [23].

Importantly, PCC is a key region in the limbic system and the DMN, and has a major role in regulating emotions and motivational thoughts [59]. Similarly, ACC which has been associated with MDD in multiple studies [8–10,13,15–18,20,21,25], has been most frequently linked to the experience of pain, and specially in monitoring painful social situations such as exclusion or rejection [60]. This may explain the implication of PCC and ACC in the high depressive network. Furthermore, the prefrontal and limbic cortical regions are directly modulated by the subcortical structures that are involved in the negative emotional circuits. For instance, the canonical amygdala, hippocampal structures, and the lateral habenula (LHb) has been found to represent key brain regions in the pathophysiology of depression [61]. Clinical and preclinical evidence implicates hyperexcitability of the lateral habenula (LHb) in the development of psychiatric disorders including MDD[62]. Hyperactive LHb strongly inhibit the dopaminergic motivational, cognitive and reward system [63] and produces depressive- and anxiety-like behaviors, anhedonia and aversion. Dopaminergic and anxiety circuits strongly modulate the limbic cortex, including the orbitofrontal cortex and the anterior cingulate cortex. The medial prefrontal cortex in rodent, the homologue of the anterior cingulate cortex of human are strongly involved in the expression of anxiety and are involved in MDD [64]. Because these internal emotional circuits strongly modulate the prefrontal, insula and limbic lobes, it is not surprising that in the high depressive network we found a dense functional connection between these areas.

According to the low depressive network, the regions implicated are SPL, LOG and PCUN. In fact, SPL, LOG and PCUN connectivity were found to be different between MDD patients and healthy controls in many studies [18,23,26]. SPL is principally involved in attention and visuomotor integration [65], whilst LOG plays a major role in the visual perception. In addition, visuospatial processing, reflections upon self, and aspects of consciousness are associated with PCUN. This may explain the implication of these regions in low depressive subjects where attentional and visual performances are more efficient.

## EEG frequency bands in depression

In this study, differences between MDD and healthy groups were only depicted in alpha band, and no significant results were obtained in other EEG frequency bands. Interestingly, using the same frequency band, the model has demonstrated its performance in predicting novel and independent samples. This finding is in line with a number of resting state EEG studies that continuously report MDD alterations in alpha frequency patterns. Particularly, alpha-asymmetry was suggested to be a potential marker for depression [26,66–68] whereas alterations in other frequency band such as delta, beta and gamma power were inconsistent [66]. More importantly, a connectivity-based EEG analysis suggests an increase in alpha-band connectivity between the anterior cingulate cortex and both the prefrontal cortex [69]. It has also been shown that reduced alpha desynchronization in a network involving bilateral frontal and right-lateralized parietal regions may provide a specific measure of deficits in approach-related motivation in depression [70]. Using graph theoretical methods, another study shows that disrupted global and local network indices in MDD patients were revealed in alpha band [11]. In addition, features derived from alpha band have been revealed to be highly discriminating when distinguishing between MDD patients and healthy controls using binary classifiers [32,34,37,71]. These observations may be explained by the fact that alpha rhythm is thought to functionally inhibit cortical responses to unattended components of sensory input [72] and monitor important roles in both cognitive, social and emotion processing [73,74] that are closely altered in MDD. In contrast, a recent study show that resting state connectivity disruptions emerged mainly in beta frequency band (12.5-21 Hz), supporting the hypothesis that beta-band synchronization corresponds to a cognitive idling rhythm present in MDD [20]. In addition, a large study including 1344 participants showed increases in source EEG theta power across frontal regions of the brain [75]. This inconsistency in the results might be due to the difference in the subjects/patients' characteristics as well the difference in methodological issues such as number of subjects, data pre-processing and data analysis (power spectra vs. connectivity for instance). Moreover, neural connectivity networks likely involve coordinated synchronizations across frequencies. Thus, extending the analysis to establish a model based on cross-frequency couplings for instance may be of interest.

## Limitations and methodological considerations

First, the model was established using the first dataset composed of two groups (healthy and MDD) where the severity of depressive symptoms was assessed using BDI score. Although the BDI is a self-report questionnaire, as compared to the HADRS, was reported that BDI revealed high reliability and good correlation with measures of depression and anxiety [76]. However, like any self-report measure, BDI suffers from intrinsic limitations. Specifically, the score can be exaggerated or minimized based on the clinical environment [77], fatigue [78], and compromised cognitive functioning [76].

Second, we are aware that subcortical regions are not easily accessible using scalp EEG, namely due to anatomical considerations. Unlike the layered cortex, a subcortical region would not have the necessary organization of pyramidal cells to give rise to localizable scalp-recorded EEG. Conversely, exploring subcortical regions may contribute to the modeling of depressive networks as many studies linked MDD with disruptions in amygdala [7,18,79], a region centrally implicated in emotional and physiologic responses.

Third, a limitation of the current study is that one of the datasets included only healthy subjects and that some subjects experiencing mild depressive symptoms. We also believe that the correlation between observed and predicted HDRS scores needs to be validated in a larger sample (including MDD patients). In addition, longitudinal data (EEG data recorded from healthy subjects and MDD patients at several time points) is mandatory to validate and ultimately generalize the potential neuromarker.

Fourth, while the low-depressive network showed significant correlation between observed and predicted BDI scores in the first dataset, the external validation didn't reveal significant correlations between observed and predicted HDRS scores in the second sample. This observation suggests that the high-depressive network showed consistent prediction of depression severity across both datasets while the low-depressive network is less relevant to the prediction performance for the second dataset. It is worth noting, however, that the first dataset provided two separated group of subjects (healthy and MDD patients), while the second didn't show a wide distribution of the HDRS score. More precisely, this latter mainly presents healthy subjects (with Hamilton scores spanning from 0 to 10) and some subjects experiencing mild depressive symptoms (Hamilton scores spanning from 10 to 14). This may explain the disability of the low-depressive network to discriminate between these subjects as it is expected to show similar low-network strength. Another explanatory possibility would be that normal sadness and pathological depression share the same continuum where major depression is defined based on the functional consequences of the symptoms (i.e. described as the pragmatic approach [80,81]). The continuum hypothesis between normal sadness and pathological major depression has also been supported at the symptom [82] and the cerebral levels [83] using data-driven latent analysis.

In conclusion, our findings demonstrate the feasibility of resting-state EEG whole-brain connectome in predicting individual differences in depression severity, suggesting that it might serve as direct and non-invasive neuromarker that can ultimately support clinicians in a biologically-based characterization of MDD, which eventually will improve MDD prognostic and therapeutic decision.

# Acknowledgments

This work was financed by the Rennes University, the Institute of Clinical Neuroscience of Rennes (Projects named EEGCog and EEGNET3). This work was also financed by the AZM and SAADE Association, Tripoli, Lebanon and by the National Council for Scientific Research (CNRS) in Lebanon. Authors would like to thank Campus France, Programme Hubert Curien CEDRE (PROJET N° 42257YA) and the Lebanese Association for Scientific Research (LASER) for their support.